\documentclass{entcs} \usepackage{entcsmacro}
\usepackage{amssymb} 
\usepackage{amsmath} 
\usepackage{latexsym} 
\usepackage{epsf} 
\usepackage[all]{xy}  
\usepackage{proof} 
\usepackage{lscape} 
\usepackage{diagrams}  
\usepackage{verbatim}    
\usepackage{times}   
\usepackage{pdfsync}  
\usepackage{color}  
\usepackage{graphicx}
\newcommand{\bpf}{\noindent {\bf Proof:} } 

\def\endproof{\hfill$\Box$}  

\def\qed{\ifmmode 
          $\Box$ 
         \else 
         {\unskip 
          \nobreak 
          \hfil 
          \penalty50 
          \hskip1em 
          \null 
          \nobreak 
          \hfil 
          $\Box$ 
          \parfillskip=0pt 
          \finalhyphendemerits=0 
          \endgraf} 
         \fi} 
 
\newcommand{\semantics}[1]{[\![ #1 ]\!]} 
\newcommand{\Gm}{\Gamma}
\newcommand{\Dl}{\Delta} 

\newcommand{\bdia}[1]{\blacklozenge_{#1}}
\newcommand{\comb}{\bullet}

\newcommand{\booktitle}[1]{``#1''}
\newcommand{\articletitle}[1]{{\emph{#1}}}
\newcommand{\thesistitle}[1]{``#1''}

\def\lastname{Sadrzadeh and Dyckhoff}
\begin{document}
\begin{frontmatter}
  \title{Positive Logic with Adjoint Modalities: \\
         Proof Theory, Semantics and 
         Reasoning about  Information} 

\author{Mehrnoosh Sadrzadeh
	\thanksref{ALL}
	\thanksref{myemail}}
\address{Computer Laboratory,\\ 
         Oxford University,\\
         Oxford, England, UK} 

\author{Roy Dyckhoff
    \thanksref{coemail}}
\address{School of Computer Science,\\
           St Andrews University,\\
           St Andrews, Scotland, UK} 
\thanks[ALL]{Support by EPSRC (grant EP/F042728/1) is gratefully acknowledged, 
	as are helpful comments by Kai Br\"unnler, Rajeev Gor\'{e}, Greg Restall and Alex Simpson.} 

\thanks[myemail]{Email:
         \href{mailto:mehrs@comlab.ox.ac.uk} 
         {\texttt{\normalshape mehrs@comlab.ox.ac.uk}}} 
\thanks[coemail]{Email:
         \href{mailto:rd@st-andrews.ac.uk} 
            {\texttt{\normalshape rd@st-andrews.ac.uk}}}

\begin{abstract}
We consider a simple modal logic whose non-modal part has conjunction and disjunction as connectives and whose modalities come in adjoint pairs, but are not in general closure operators. Despite absence of negation and implication, and of axioms corresponding to the characteristic axioms of (e.g.) T, S4 and S5, such logics are useful, as shown in previous work by Baltag, Coecke and the first author, for encoding and reasoning about information and misinformation in multi-agent systems. For such a logic we present an algebraic semantics, using lattices with agent-indexed families of adjoint pairs of operators, and a cut-free sequent calculus. The calculus exploits operators on sequents, in the style of ``nested'' or ``tree-sequent'' calculi; cut-admissibility is shown by constructive syntactic methods. The applicability of the logic is illustrated by reasoning about the muddy children puzzle, for which the calculus is augmented with extra rules to express the facts of the muddy children scenario. 
\end{abstract}

\end{frontmatter}

\section{Introduction}
Modal logics include various modalities, represented as unary operators, used to formalize and reason about extra modes such as time, provability, belief and knowledge, applicable in various areas (we have that of security protocols in mind). Like disjunction and conjunction, modalities often come in pairs, e.g.\ $\Diamond$ and $\Box$: one preserves disjunctions and the other conjunctions.  According to the intended application, further axioms such as monotonicity and idempotence can be imposed on the modalities. 

As well as relational (or Kripke) models, one may consider as models for such logics various ordered structures, such as lattices with operators. The question then arises as to what is the simplest way of obtaining a pair of these operators. If the lattice is a Boolean Algebra and thus has negation, any join-preserving operator (such as $\Diamond$) immediately provides us with a meet-preserving one (such as $\Box$) by de Morgan duality. In a Heyting Algebra, the lack of De Morgan duality will cause one of these operators to preserve meets only in one direction.  What if no negation is present, e.g. in a distributive lattice? The categorical notion of adjunction (aka Galois connection) is useful here: any (arbitrary) join-preserving endomorphism on a lattice has a Galois right adjoint, which (by construction) preserves meets. For example, in category {\tt Sup} of sup-lattices with join-preserving maps, every such map is residuated, i.e. has a right adjoint.    

In this paper, we consider a minimal modal logic where the underlying logic has only two binary non-modal connectives---conjunction and disjunction---and where the modalities are adjoint but have no closure-type properties (such as idempotence). As algebraic semantics one may consider a bounded distributed lattice, the modalities thereof being residuated lattice endomorphisms. Examples are quantales and Heyting algebras when one argument of their residuated multiplication and conjunction (respectively) is fixed.  One may also consider a relational semantics. In the proof of relational completeness, the absence of negation prohibits us from following standard canonical model constructions, as we can no more form maximally consistent sets. We overcome this by developing an equivalent Hilbert-style axiomatization for our logic and then using the general Sahlqvist results of~\cite{GehrkeNagahashiVenema} based on completion of  algebras with operators.  

We provide a sequent calculus, which contains, in addition to axioms for the logical constants $\top, \bot$ only the operational  left and right rules for each connective and operator. We prove admissibility of the structural rules of {\em Contraction}, {\em Weakening} and {\em Cut}  by constructive syntactic methods.  In the absence of negation and of closure-type properties for the modalities, developing well-behaved sequent calculus rules for the modalities (in particular the left rule for the right adjoint $\Box$) was a challenging task; a calculus not obviously allowing cut-elimination was given in \cite{BaltagCoeckeSadr}.  Our sequents are a  generalization of Gentzen's  where the contexts (antecedents of sequents), as well as formulae, have a structure and can be nested.  
For application, we augment our calculus  with a rule that allows us to encode assumptions of epistemic scenarios, and show that {\em Cut} is still admissible. 

We interpret our adjoint modalities as \emph{information} and \emph{uncertainty} and use them to encode and prove epistemic properties of the puzzle of muddy children. Due to the absence  of negation, we can only express and prove positive versions of these epistemic properties. But, our proofs are simpler  than the proofs of traditional modal logics, e.g. those in~\cite{Huth}. In a nutshell,   in just one proof step the adjunction is unfolded  and   the information modality is replaced by the  \emph{uncertainty} modality; in the next proof step, the assumptions of the scenario  are imported  into the logic via the assumption rule. At this stage the modalities  are eliminated and the proof continues in a propositional setting. Since our information modality is not necessarily truthful, we are able to  reason about more challenging versions of  epistemic scenarios, for example when  agents are dishonest and their deceitful communications lead to false information. Properties of these more challenging versions have not been proved in traditional modal logic in computer science approaches, like that of~\cite{Huth}.

From  a proof theoretic point of view, it is the first time that a cut-free sequent calculus has been developed  for logics where the underlying propositional logic has no negation and implication and where  the modalities are adjoint. There is a gap in the literature on proof theory for modal logic for the combination of the two.  The calculi specifically designed to deal with adjoint connectives, such as the display calculi~\cite{Belnap,Gore,Moortgat} need  to have at least one adjoint binary operator and face a challenge in developing cut-free rules for our positive non-adjoint  connectives. Perhaps the closest to our proof systems are deep inference systems; but the modal deep inference systems developed so far, e.g. by Brunnler~\cite{Brunnler} and Kashima~\cite{Kashima} include negation and implication.  Of these two formalisms, the closest to ours is that for the tense logic of Kashima.  Other than differences in logic (presence of negation and two-sided sequents), which lead to different modal rules (based on de Morgan duality), our proof theoretic techniques have a merit over  those of Kashima:  (1) we formalize deep substitution in the nested sequents and as a result do not need to develop two different versions of the calculus and prove soundness  and completeness separately, and (2)  our cut-admissibility proofs are done explicitly via a syntactic construction rather than as a consequence of semantic completeness.  

Some proof systems encode modalities  by introducing semantic labels to encode accessibility and satisfaction relations; these are better placed to produce cut-free systems for adjoint modalities, e.g. the comprehensive work of  Negri~\cite{Negri} and ~\cite{Simpson} for classical and intuitionistic modal logics.  The former in particular can be adapted to provide a cut-free labelled sequent calculus for our logic. However, that these systems are strongly based on the relational semantics of modal logic and mix it with the syntax of the logic does not fit well with the spirit of the algebraic motivation behind our minimal logic. 

On the application side, adjoint modalities have been originally used to reason about time in the context of tense logics,  e.g. in~\cite{Prior,vonKarger}. Their epistemic application is novel and was initiated in the dynamic epistemic algebra of~\cite{BaltagCoeckeSadr,SadrThesis}, constituting a pair of an endowed quantale of ``actions'' and its right module of ``propositions'' and abstracting over  the Dynamic Epistemic Logic of~\cite{BaltagMoss}.  Our work provides an answer to two open problems in that work:  (1) elimination of the cut rule for [a variant of] the calculus corresponding to the propositional part of the algebra, and (2) complete relational models and representation theorems for the propositional part.

\section{Sequent calculus for positive logic with adjoint modalities}
\subsection{Sequent Calculus} 
We refer to our logic as $APML$ for ``adjoint positive modal logic'', with the suffix $Tree$ when we consider a tree-style sequent calculus. The set $M$ of {\em formulae} $m$ of our logic is generated over a set ${\cal A}$ of {\em agents} $A$ and a set $At$ of {\em atoms} $p$  by the following grammar:
$$ m ::= \bot \mid \top \mid p  \mid m \wedge m \mid m \vee m  \mid   
\Box_A\,m \mid \bdia{A}(m)
$$

{\em Items} $I$ and {\em contexts} $\Gm$ are generated by the following syntax:
$$		I		::=		m 	\mid \Gm^A
	\quad\quad
		\Gamma 	::=  I\ multiset
$$
where $\Gm^A$ will be interpreted as $\bdia{A}(\bigwedge \Gm)$, for $\bigwedge \Gm$ the conjunction of the interpretations of elements in  $\Gm$. 

Thus, {\em contexts} are finite multi-sets of items, whereas {\em items} are either formulae or agent-annotated contexts. The use of multi-sets rather than sets makes the role of the {\em Contraction} rule explicit, with the rules in a form close to the requirements of an implementation. The union of two multi-sets is indicated by a comma, as in $\Gm, \Gm'$ or (treating an item $I$ as a one element multiset) as in $\Gm, I$. 

If one of the items inside a context is replaced by a ``hole'' $[]$, we have a {\em context-with-a-hole}. More precisely, we have the notions of  {\em context-with-a-hole} $\Dl$ and {\em item-with-a-hole} $J$, defined using mutual recursion as follows:
$$
\Dl 	::=  \Gm, J \quad\quad\quad
J      ::=   []  \mid \Dl^A
$$
\noindent
and so a context-with-a-hole is a context (i.e.\ a multiset of items) together with an {\em item-with-a-hole}, i.e.\ either a hole or an agent-annotated context-with-a-hole.  To emphasise that a context-with-a-hole is not a context, we use $\Dl$ for the former and $\Gm$ for the latter; similarly for items-with-a-hole $J$ and items $I$.

Given a context-with-a-hole $\Dl$ and a context $\Gm$, the result $\Dl[\Gm]$ of applying the first to the second, i.e.\ replacing the hole $[]$ in $\Dl$ by $\Gm$, is a context, defined recursively (together with the application of an item-with-a-hole to a context) as follows:
$$
(\Gm', J) [\Gm] 	 	=	 \Gm', J[\Gm] 	\quad\quad\quad
([]) [\Gm] 		 	=	 \Gm 			\quad\quad
(\Dl^A) [\Gm]  	 	=	 \Dl[\Gm]^A
$$
\noindent

Given contexts-with-a-hole $\Dl', \Dl$, and an item-with-a-hole $J$, the {\em combinations} $\Dl' \comb \Dl$ and $J \comb \Dl$ are defined as follows by mutual recursion on the structures of $\Dl'$ and $J$:
$$
(\Gm, J) \comb \Dl 	 	=	 \Gm, ( J \comb \Dl) 	\quad\quad\quad
([]) \comb \Dl 		 	=	 \Dl 			\quad\quad
(\Dl''^A) \comb \Dl  	 	=	 (\Dl''\comb \Dl)^A
$$

\begin{lemma} \label{comb-lemma} Given contexts-with-a-hole $\Dl', \Dl$, an item-with-a-hole $J$ and a context $\Gm$, the following hold:
$$
(\Dl' \comb \Dl) [\Gm] 	 	=	\Dl'[\Dl[\Gm]]	\quad\quad
   (J \comb \Dl) [\Gm] 	 	=	J[\Dl[\Gm]]
$$
\end{lemma}
\bpf Routine. 
\qed

We have the following initial sequents (in which $p$ is restricted to being an atom):
\begin{center}
\fbox{$
\begin{array}{ccc}
\infer[Id]{\Gm,p \vdash p}{} \qquad & \qquad  
\infer[\bot L]{\Dl[\bot] \vdash m}{} \qquad & \qquad 
\infer[\top R]{\Gm \vdash \top}{}
\end{array}
$}
\end{center}

\medskip \noindent
The rules for the lattice operations and the modal operators are: 
\begin{center}
\fbox{$
\begin{array}{cc}
\infer[\wedge L]
	{\Dl [m_1 \wedge m_2] \vdash m}
		{\Dl [m_1,  m_2] \vdash m} 
&	 
\infer[\wedge R]
	{\Gm \vdash m_1 \wedge m_2}
		{\Gm \vdash m_1 \quad \Gm \vdash m_2} 
\\ \ \\
\infer[\vee L]
	{\Dl [m_1 \vee m_2] \vdash m}
		{\Dl[m_1] \vdash m \quad \Dl [m_2] \vdash m}
&
\qquad 
\infer[\vee R1]
		{\Gm \vdash m_1 \vee m_2}
			{\Gm \vdash m_1}
\qquad 
\infer[\vee R2]
		{\Gm \vdash m_1 \vee m_2}
			{\Gm \vdash m_2}
\\ \ \\
\infer[\bdia{A} L]
	{\Dl [\bdia{A}(m)] \vdash m'}
		{\Dl [m^A] \vdash m'}
&
\infer[\bdia{A} R]
	{\Gm', \Gm^A \vdash \bdia{A}(m)}
		{\Gm \vdash m} 
\\ \ \\
\infer[\Box_A L]
	{\Dl [(\Box_A m, \Gm)^A] \vdash m'}
		{\Dl [(\Box_A m, \Gm)^A, m] \vdash m'}
&
\infer[\Box_A R]
	{\Gm \vdash \Box_A\,m}
		{\Gm^A \vdash  m}
\end{array}
$}
\end{center}

The two indicated occurrences of $p$ in the {\em Id} rule are {\em principal}.
Each right rule has its conclusion's succedent as its {\em principal formula}; in addition, the $\bdia{A} R$ rule has $\Gm^A$ as a {\em principal item} and $\Gm'$ (which is there to ensure admissibility of {\em Weakening}) as its {\em parameter}.
Each left rule has a {\em principal item}; these are as usual, except that the $\Box_A L$ rule has  the formula $\Box_A m$ {\em principal} as well as the principal item $(\Box_A m, \Gm)^A$. 

Note that the $\Box_A L$ rule duplicates the principal item in the conclusion into the premiss; in examples, we may omit this duplicated item for simplicity. This duplication allows a proof of the admissibility of {\em Contraction}, and thus of completeness. To see its necessity, note that the following sequent is (according to the algebraic semantics in Section \ref{semantics}) valid:
$$\bdia{A}(\Box_A(m \vee n) )  \leq (m \wedge \bdia{A}(\Box_A(m \vee n))) \vee ( n \wedge \bdia{A}(\Box_A(m \vee n)))$$

\noindent
It is, however, not derivable unless the principal item of $\Box_A L$ is duplicated into the rule's premiss.

As a standard check on the rules, we show the following:
\begin{lemma}\label{id}
For every formula $m$ and every context $\Gm$, the sequent $\Gm, m \vdash m$ is derivable.
\end{lemma}
\bpf By induction on the size of $m$. In case $m$ is an atom, or $\bot$, or $\top$, the sequent $\Gm, m \vdash m$ is already initial. For compound $m$, consider the cases. Meet and join are routine. 
 Suppose $m$ is $\bdia{A}(m')$; by inductive hypothesis, we can derive $m'  \vdash m'$, and by $\bdia{A}R$ we can derive $\Gm, m'^A \vdash \bdia{A}(m')$, whence $\Gm, \bdia{A}(m') \vdash \bdia{A}(m')$ by $\bdia{A} L$. 

Now suppose $m$ is $\Box_{A} m'$.
By inductive hypothesis, we can derive $(\Gm, \Box_A m')^A, m' \vdash m'$, and by $\Box_A L$ we get $(\Gm, \Box_A m')^A \vdash m'$; from this we obtain $\Gm, m \vdash m$ by $\Box_A R$. \qed
\vskip 6pt

Since we use multisets (for contexts) rather than sets or lists, the rules of exchange and associativity are inexpressible.

To allow induction on the sizes of items, we need a precise definition, with a similar definition for contexts. The {\em size of a formula} is just the (weighted) number of operator occurrences, counting each operator $\bdia{A}$ and $\Box_A$ as having weight $2$; the {\em size of an item} $\Gm^A$ is the size of $\Gm$ plus $1$, and the {\em size of a context} is the sum of the sizes of its items. The {\em size of a sequent} $\Gm \vdash m$ is just the sum of the sizes of $\Gm$ and $m$.  Note that each premiss of a rule instance has lower size than the conclusion, except for the rule $\Box_A L$, whose presence leads to non-termination of a naive implementation of the calculus. 

\begin{lemma}\label{weak-lab}
The following {\em Weakening} rule  is admissible:
$$
	\infer[Wk]{\Dl[\Gm, \Gm'] \vdash m}{\Dl[\Gm] \vdash m}
$$
\end{lemma}
\bpf Induction on the depth of the derivation of the premiss and case analysis (on the rule used in the last step). 
Suppose the last step is by $\bdia{A} R$, with $m = \bdia{A} m'$, and with premiss $\Gm^{*} \vdash m'$ and conclusion $\Gm'',\Gm^{*A} \vdash m$, so $\Dl[\Gm] = \Gm'',\Gm^{*A}$. To obtain $\Dl[\Gm, \Gm']$ from this there are two possibilities. In the first case, $\Gm$ occurs inside $\Gm^{*A}$, and we make a routine use of the inductive hypothesis and reapply $\bdia{A} R$  with the same  parameter. In the second case, we just use the $\bdia{A} R$ rule with a different parameter. Other cases are straightforward. 
\qed

\begin{lemma}\label{invert-lab}
The $\bdia{A} L$ and $\Box_A R$ rules are invertible, i.e. the following are admissible:
$$
	\infer[\bdia{A} Inv]{\Dl [m^A] \vdash m'}{\Dl [\bdia{A}(m)] \vdash m'}
\qquad \qquad
	\infer[\Box_A Inv]{\Gm^A \vdash m}{\Gm \vdash \Box_A m}
$$
\end{lemma}
\bpf
Induction on the height of the derivation of the premiss.  \qed

\begin{lemma}\label{invert-lab2}
The $\wedge L$, $\vee L$ and $\wedge R$ rules are invertible.
\end{lemma}
\bpf
Induction on the height of the derivation of the premiss. \qed

\begin{lemma} \label{Icontr}
The following {\em Item Contraction}  rule is admissible
$$
\infer[IContr]{\Dl[ I] \vdash m}{\Dl[I,I] \vdash m}
$$
\end{lemma}
\bpf Strong induction on the size of the item $I$, with a subsidiary induction on the height of the derivation of the premiss, together with case analysis and the above inversion lemmas. Consider the cases of the last step; first, when $I$ is non-principal, we permute the contraction up and (keeping $I$ fixed) apply the subsidiary induction hypothesis; when the premiss is an initial sequent, so is the conclusion; when the step is by $\bdia{A} R$ with $I$ principal (and thus of the form $\Gm'^A$) the premiss of that step has antecedent $\Gm'$ from which the copy of $I$ is absent, allowing reuse of the $\bdia{A} R$ rule to yield $\Dl[I] \vdash m$;  and, when $I$ is otherwise principal, the last step is one of the four one-premiss left rules. The $\Box_A L$ case is handled by the subsidiary inductive hypothesis (for the two cases, where $I$ is an item $(\Box_A m', \Gm)^A$ and where it is a formula $\Box_A m'$ inside such an item), and the other cases ($\wedge L, \vee L, \bdia{A} L$) are handled by the invertibility lemmas and the main inductive hypothesis. \qed

\begin{corollary}
The following {\em Contraction}  rule is admissible
$$
\infer[Contr]{\Dl[ \Gamma] \vdash m}{\Dl[\Gamma,\Gamma] \vdash m}
$$
\end{corollary}
\bpf Induction on the size of the context $\Gamma$, by Lemma \ref{Icontr}. \qed

\begin{lemma}\label{topL-Sub-weak}
The rule $\top L^{-}$  is admissible:
$$\infer[\top L^{-}]{\Dl[\Gm] \vdash m}{\Dl[\top] \vdash m}
$$
\end{lemma}
\bpf Induction on the depth of the derivation of the premiss and case analysis. \qed

\begin{theorem}
The {\em Cut}   rule is admissible
$$
	\infer[Cut]{\Dl' [\Gm] \vdash m'}{\Gm \vdash m \quad \Dl' [m] \vdash m'}
$$
\end{theorem}
\bpf
Strong induction on the rank of the cut, where the {\em rank} is given by the pair (size of cut formula $m$, sum of heights of derivations of premisses).

To clarify the different reductions used (and to show how all cases are covered), we present the different cases in tabular form: in the top row are the different cases for the last step of the first premiss of the cut and in the left column are the different cases for the last step of the derivation of the second premiss of the cut. The letters refer to the case in the treatment above; where there are two letters, either reduction may be used. The attributes like ``Non-Principal'' refer to the status of the cut formula w.r.t.\ the rule. The proof for each case is presented in the appendix.

\bigskip
\newcommand{\np}{Non-Principal}
\newcommand{\pr}{Principal}
\newcommand{\na}{Non-Auxiliary}
\newcommand{\au}{Auxiliary}
		
\begin{tabular}{|l| l l l l l l l l l l l|}
\hline
\quad 								&Id &$\bot L$ & $\top R$ & $\wedge L$ & $\vee L$ & $f_A L$ & $\Box_A L$ &
															  $\wedge R$ & $\vee R$ & $f_A R$ & $\Box_A R$\\
\hline
$Id$ (\pr) &					(i) & (ii) & (iii) & (iv) & (v) & (vi) & (vii) & (viii) & (ix) & (x) &  (xi) (a)\\
$\bot L$ (\np)& 				$\downarrow$ &$\downarrow$ &$\downarrow$ &$\downarrow$ &$\downarrow$ &$\downarrow$ &
							$\downarrow$ &$\downarrow$ &$\downarrow$ &$\downarrow$ & (xi) (b)\\
$\top R$& 					&&&&&&&&&& (xi) (c)\\

\hline						
$\wedge L$ (\np)& 			&&&&&&&&&& (xi) (d)\\
$\vee L$ (\np)& 				&&&&&&&&&& (xi) (e) \\
$f_B  L$ (\np)& 				&&&&&&&&&& (xi) (f)\\
$\Box_B  L$ (\np)& 			&&&&&&&&&& (xi) (g)\\

\hline
$\Box_A  L$ (\pr)& 			&&&&&&&&&& (xi) (h)\\

\hline
$\wedge R$& 					&&&&&&&&&& (xi) (i)\\
$\vee R$& 					&&&&&&&&&& (xi) (j)\\
$f_B R$& 					&&&&&&&&&& (xi) (k)\\
$\Box_B R$& 					&&&&&&&&&& (xi) (l)\\

\hline
\end{tabular} 

    \begin{enumerate} 
\item	The first premiss is an instance of $Id$. 
		$$
									\infer[Cut]{\Dl' [\Gm', p] \vdash m'}{
											\infer[Id]{\Gm', p \vdash p}{} 
											& 
											\Dl' [p] \vdash m'
										}
										$$
								is transformed to
									$$\infer[Wk]
										{\Dl' [\Gm', p] \vdash m'}
											{\Dl' [p] \vdash m'}
									$$
\item	The first premiss is an instance of $\bot L$.  $$
									\infer[Cut]{\Dl' [\Dl [\bot]] \vdash m'}{
									\infer[\bot L]{\Dl [\bot] \vdash p}{}
									&
									\Dl' [p] \vdash m'
									}
									$$
								is transformed 
								(using Lemma \ref{comb-lemma} to identify 
								$\Dl'[\Dl[\bot]] = (\Dl' \comb \Dl) [\bot]$)
								to
									$$
									\infer[\bot L]{\Dl' [\Dl[\bot]] \vdash m'}{}
									$$ 

\item	The first premiss is an instance of $\top R$.
							$$
							\infer[Cut]{\Dl'[\Gm] \vdash m'}{
								\infer[\top R]{\Gm \vdash \top}{}
								&
								\Dl' [\top] \vdash m'}
							$$
							 transforms to the following using Lemma \ref{topL-Sub-weak}
							$$
							\infer[\top L^{-}]{\Dl' [\Gm] \vdash m'}{
								{\Dl' [\top] \vdash m'}}
							$$
\item	The first premiss is an instance of $\wedge L$.    Straightforward
\item	The first premiss is an instance of $\vee L$.		Straightforward
\item	The first premiss is an instance of $\bdia{A} L$. 
				$$
				\infer[Cut]{\Dl' [\Dl [\bdia{A}(m)]] \vdash m''}{
				\infer[\bdia{A} L]{\Dl [\bdia{A}(m)] \vdash m'}{\Dl [m^A] \vdash m'} 
				&
				\Dl' [m'] \vdash m''
				}
				$$
				transforms (using Lemma \ref{comb-lemma}) to
				$$
				\infer[\bdia{A} L]{\Dl' [\Dl [\bdia{A}(m)]] \vdash m''}{
				\infer[Cut]{\Dl' [\Dl [m^A]] \vdash m''}
				{\Dl [m^A] \vdash m' \qquad \Dl' [m'] \vdash m''}
				}
				$$
\item	The first premiss is an instance of $\Box_A L$.
				$$
				\infer[Cut]{\Dl' [\Dl [(\Box_A m, \Gm)^A]] \vdash m''}{
				\infer[\Box_A L]{\Dl [(\Box_A m, \Gm)^A] \vdash m'}{\Dl [(\Box_A m, \Gm)^A,m] \vdash m'}
				&
				\Dl'[m'] \vdash m''}
				$$
				transforms  to 
				$$
				\infer[\Box_A L]{\Dl' [\Dl [(\Box_A m, \Gm)^A]] \vdash m''}
					{\infer[Cut]{\Dl' [\Dl [(\Box_A m, \Gm)^A, m]] \vdash m''}
						{\Dl [(\Box_A m, \Gm)^A,m] \vdash m' \qquad \Dl'[m'] \vdash m''}
				}
				$$
\item	The first premiss is an instance of $\wedge R$.   
				$$
					\infer[Cut]{\Dl[\Gm] \vdash m'}
						{
						\infer[\wedge R]{\Gm \vdash m_1 \wedge m_2}{\Gm \vdash m_1 \qquad \Gm \vdash m_2}
						&
						\Dl[m_1 \wedge m_2] \vdash m'
						}
					$$
					transforms to
					$$
					\infer[Contr]{\Dl[\Gm] \vdash m'}
						{\infer[Cut]{\Dl[\Gm, \Gm] \vdash m'}
							{\Gm \vdash m_2 
								&
							\infer[Cut]{\Dl[\Gm, m_2] \vdash m'}{\Gm \vdash m_1 &
							\infer[Inv \wedge L]{\Dl[m_1, m_2] \vdash m'}{\Dl[m_1 \wedge m_2] \vdash m'}}
							}
						}
					$$

\item	The first premiss is an instance of $\vee R$.
			$$
				\infer[Cut]{\Dl[\Gm] \vdash m'}
					{
					\infer[\vee R1]{\Gm \vdash m_1 \vee m_2}{\Gm \vdash m_1}
					&
						\Dl[m_1 \vee m_2] \vdash m'
					}
				$$
					transforms to
				$$
				\infer[Cut]{\Dl[\Gm] \vdash m'}
					{\Gm \vdash m_1
					&
					\infer[Inv \vee L]{\Dl[m_1] \vdash m'}
						{\Dl[m_1 \vee m_1] \vdash m'}
					}					
				$$

\item	The first premiss is an instance of $\bdia{A} R$.
						$$
				   \infer[Cut]{\Dl'[\Gm', \Gm^A] \vdash m'}{
				   	\infer[\bdia{A} R]{\Gm', \Gm^A \vdash \bdia{A}(m)}{\Gm \vdash m}
				   	&
					 \Dl' [\bdia{A}(m)] \vdash m'
					}
				   $$
				   is transformed to
				   $$
				   \infer[Wk]{\Dl'[\Gm', \Gm^A] \vdash m'}
				   	{\infer[Cut]{\Dl'[\Gm^A] \vdash m'}
				   		{\Gm \vdash m 
						&
						\infer[Inv \bdia{A} L]{\Dl'[m^A] \vdash m'}
							{\Dl'[\bdia{A}(m)] \vdash m'
							}
						}
					}
				   $$

\item	The first premiss is an instance of $\Box_A R$. This now depends on the form of the second premiss. 
	\begin{enumerate}
	\item Id
	$$
	\infer[Cut]{\Dl[\Gm], p \vdash p}{
		\infer[\Box_A R]{\Gm \vdash \Box_A m}{\Gm^A \vdash m
		}
		&
		\infer[Id]{\Dl[\Box_A m], p \vdash p}{}
		}
	$$
	transforms to
	$$
	\infer[Id]{\Dl[\Gm], p \vdash p}{}
	$$
	\item $\bot L$
	$$
	\infer[Cut]{\Dl[\Gm][\bot] \vdash m'}
		{
		\infer[\Box_A R]{\Gm \vdash \Box_A m}{\Gm^A \vdash m}
		&
		\infer[\bot L]{\Dl[\Box_A m][\bot] \vdash m'}{}
		}
	$$
	transforms to
	$$
	\infer[\bot L]{\Dl[\Gm][\bot]\vdash m'}{}
	$$
	\item $\top R$
	$$\infer[Cut]{\Dl[\Gm] \vdash \top}{
						\infer[\Box_A R]{\Gm \vdash \Box_A m}{\Gm^A \vdash m}
						&
						\infer[\top R]{\Dl[\Box_A m] \vdash \top}{}
						}
					$$
					transforms to
					$$
					\infer[\top R]{\Dl[\Gm] \vdash \top}{}
					$$

	\item $\wedge L$, non-principal
	$$
	\infer[Cut]{\Dl[\Gm][m_1 \wedge m_2] \vdash m'}
		{
		\infer[\Box_A R]{\Gm \vdash \Box_A m}{\Gm^A \vdash m}
		&
		\infer[\wedge L]{\Dl[\Box_Am][m_1 \wedge m_2] \vdash m'}{\Dl[\Box_Am][m_1, m_2] \vdash m'}
		}
	$$
	transforms to
	$$
	\infer[\wedge L]{\Dl[\Gm][m_1 \wedge m_2] \vdash m'}
		{\infer[Cut]{\Dl[\Gm][m_1, m_2] \vdash m'}
			{\Gm \vdash \Box_A m \qquad \Dl[\Box_A m][m_1, m_2] \vdash m'}
		}
	$$
	\item $\vee L$, non-principal
	$$
	\infer[Cut]{\Dl[\Gm][m_1 \vee m_2] \vdash m'}
		{\infer[\Box_A R]{\Gm \vdash \Box_A m}{\Gm^A \vdash m}
			&
		\infer[\vee L]{\Dl[\Box_A m][m_1 \vee m_2] \vdash m'}
			{\Dl[\Box_A m][m_1] \vdash m' 
				&
			\Dl[\Box_A m][m_2] \vdash m'
			}
		}
	$$
	transforms to
	$$
	\infer[\vee L]{\Dl[\Gm][m_1 \vee m_2] \vdash m'}
		{\infer[Cut]{\Dl[\Gm][m_1] \vdash m'}{\Gm \vdash \Box_A m \qquad \Dl[\Box_A m][m_1] \vdash m'}
		&
		\infer[Cut]{\Dl[\Gm][m_2] \vdash m'}{\Gm \vdash \Box_A m \qquad \Dl[\Box_A m][m_2] \vdash m'}
		}
	$$
   \item $\bdia{B} L$, non-principal
	$$
	\infer[Cut]{\Dl[\Gm][\bdia{B}(m'')] \vdash m'}
		{\infer[\Box_A R]
				{\Gm \vdash \Box_A m}
					{\Gm^A \vdash m}
		&
		\infer[\bdia{B} L]
				{\Dl[\Box_A m][\bdia{B}(m'')] \vdash m'}
					{\Dl[\Box_A m][[m'']^B] \vdash m'}
		}
	$$
	transforms to
	$$
	\infer[\bdia{B} L]
		{\Dl[\Gm][\bdia{B}(m'')] \vdash m'}
		{
		\infer[Cut]
			{\Dl[\Gm][[m'']^B] \vdash m'}
			{\Gm \vdash \Box_A m 
			&
			 \Dl[\Box_A m][[m'']^B] \vdash m'
			}
		}
	$$
	\item $\Box_B L$, non-principal
	$$
	\infer[Cut]{\Dl[\Gm][(\Box_B m'', \Gm')^B] \vdash m'}
			{\infer[\Box_A R]{\Gm \vdash \Box_A m}{\Gm^A \vdash m}
			&
			\infer[\Box_B L]{\Dl[\Box_A m][(\Box_B m'', \Gm')^B] \vdash m'}
				{\Dl[\Box_A m][(\Box_B m'', \Gm')^B, m''] \vdash m'}
			}
	$$
	transforms to
	$$
	\infer[\Box_B L]{\Dl[\Gm][(\Box_B m'', \Gm')^B] \vdash m'}
		{\infer[Cut]{\Dl[\Gm][(\Box_B m'', \Gm')^B, m''] \vdash m'}
			{\Gm \vdash \Box_A m \qquad \Dl[\Box_A m][(\Box_B m'', \Gm')^B, m''] \vdash m'}
		}
	$$
   \item $\Box_A L$, principal
	$$
					\infer[Cut]
						{\Dl' [(\Gm,\Gm')^A] \vdash m'}
							{\infer[\Box_A R]
								{\Gm \vdash \Box_A m}
									{\Gm^A \vdash m}
							&
							\infer[\Box_A L]
								{\Dl'[(\Box_A m, \Gm')^A] \vdash m'}
								{\Dl'[(\Box_A m, \Gm')^A, m] \vdash m'}
							}
					$$
					transforms to
					$$
					\infer[Contr]
						{\Dl'[(\Gm, \Gm')^A] \vdash m}
						{\infer[Wk]
							{\Dl'[(\Gm, \Gm')^A,(\Gm, \Gm')^A] \vdash m'}
								{\infer[Cut]
									{\Dl'[(\Gm, \Gm')^A, \Gm^A] \vdash m'}
										{\Gm^A \vdash m
										&
									 	\infer[Cut]
											{\Dl'[(\Gm, \Gm')^A, m] \vdash m' }
												{\Gm \vdash \Box_A m 
							 					\qquad 
							 					\Dl'[(\Box_A m, \Gm')^A, m] \vdash m'
												}
										}
								}
							}
					$$
	\item $\wedge R$
	$$
	\infer[Cut]{\Dl[\Gm] \vdash m_1 \wedge m_2}
		{\infer[\Box_A R]{\Gm \vdash \Box_A m}{\Gm^A \vdash m}
		&
		\infer[\wedge R]{\Dl[\Box_A m] \vdash m_1 \wedge m_2}
			{\Dl[\Box_A m] \vdash m_1 \qquad \Dl[\Box_A m] \vdash m_2}
		}
	$$
	transforms to
	$$
	\infer[\wedge R]{\Dl[\Gm] \vdash m_1 \wedge m_2}
		{
		\infer[Cut]{\Dl[\Gm] \vdash m_1}{\Gm \vdash \Box_A m \qquad \Dl[\Box_A m] \vdash m_1}
		&
		\infer[Cut]{\Dl[\Gm] \vdash m_2}{\Gm \vdash \Box_A m \qquad \Dl[\Box_A m] \vdash m_2}
		}
	$$
	\item $\vee R$
	$$
	\infer[Cut]{\Dl[\Gm] \vdash m_1 \vee m_2}
		{\infer[\Box_A R]{\Gm \vdash \Box_A m}{\Gm^A \vdash m}
		&
		\infer[\vee R_i]{\Dl[\Box_A m] \vdash m_1 \vee m_2}{\Dl[\Box_A m] \vdash m_i}
		}
	$$
	transforms to
	$$
	\infer[\vee R_i]{\Dl[\Gm] \vdash m_1 \vee m_2}
		{\infer[Cut]{\Dl[\Gm] \vdash m_i}{\Gm \vdash \Box_A m \qquad \Dl[\Box_A m] \vdash m_i}
		}
	$$
	\item $\bdia{B} R$
	$$
	\infer[Cut]{\Gm', \Dl[\Gm]^B \vdash \bdia{B}(m')}
		{\infer[\Box_A R]{\Gm \vdash \Box_A m}{\Gm^A \vdash m}
		&
		\infer[\bdia{B} R]{\Gm', \Dl[\Box_A m]^B \vdash \bdia{B}(m')}{\Dl[\Box_A m] \vdash m'}
		}
	$$
	transforms to
	$$
	\infer[\bdia{B} R]{\Gm', \Dl[\Gm]^B \vdash \bdia{B}(m')}
		{\infer[Cut]{\Dl[\Gm] \vdash m'}{\Gm \vdash \Box_A m \qquad \Dl[\Box_A m] \vdash m'}
		}
	$$
	\item $\Box_B R$
	$$
	\infer[Cut]{\Dl[\Gm] \vdash \Box_B m'}
		{\infer[\Box_A R]{\Gm \vdash \Box_A m}{\Gm^A \vdash m}
		&
		\infer[\Box_B R]{\Dl[\Box_A m] \vdash \Box_B m'}{\Dl[\Box_A m]^B \vdash m'}
		}
	$$
	transforms to
	$$
	\infer[\Box_B m']{\Dl[\Gm] \vdash \Box_B m'}
		{\infer[Cut]{\Dl[\Gm]^B \vdash m'}
			{\Gm \vdash \Box_A m \qquad \Dl[\Box_A m]^B \vdash m'}
		}
	$$
	\end{enumerate}
\end{enumerate}




\begin{lemma}\label{K}
The following rule (the name $K$ is roughly from \cite{Moortgat}) is admissible:
\[
\infer[K]
	{\Dl[(\Gm, \Gm')^A] \vdash m}
		{\Dl[\Gm^A, \Gm'^A, (\Gm, \Gm')^A] \vdash m}
\]
\end{lemma}
\bpf Let $\gamma = \bigwedge \Gm$ and $\gamma' = \bigwedge \Gm'$. The proof uses $Cut$ and is as follows, where a superfix $*$ indicates several instances of a rule:
$$
\infer[Contr^*]{\Dl[(\Gm, \Gm')^A] \vdash m}
	{\infer[Cut^*]{\Dl[(\Gm, \Gm')^A, (\Gm, \Gm')^A, (\Gm, \Gm')^A] \vdash m}
		{ \infer[\wedge R^*]{\Gm, \Gm' \vdash \gamma}{\dots}
		 &
		  \infer[\wedge R^*]{\Gm, \Gm' \vdash \gamma'}{\dots}
		 & 
		   \infer[\wedge L ^*]
			 {\Dl[\gamma^A, \gamma'^A, (\Gm, \Gm')^A] \vdash m}
				{\Dl[\Gm^A, \Gm'^A, (\Gm, \Gm')^A] \vdash m}
		}
	}
$$

\section{Semantics}\label{semantics}

\subsection{Algebraic Semantics}

\begin{definition}
Let $\cal A$  be  a set, with elements called {\em agents}. A {\tt DLAM} over $\cal A$ is a bounded   distributive lattice $(L, \top, \bot)$ with two $\cal A$-indexed families  $\{\bdia{A}\}_{A \in {\cal A}} \colon  L \to L$ and $\{\Box_A\}_{A \in {\cal A}} \colon  L \to L$ of order-preserving maps, with each $\bdia{A}$ left adjoint to $\Box_A$. Thus, the following hold, for all $l, l' \in L$:
\begin{eqnarray}
 l \leq l' &	\text{implies} & \bdia{A}(l) \leq \bdia{A}(l')\\
 l \leq l' &	\text{implies} & \Box_A(l) \leq \Box_A(l')\\
 \bdia{A}(l) \leq l'  & \text{iff}& l \leq \Box_A(l')
 \end{eqnarray}
\end{definition}

\begin{proposition}\label{weakdist}
In any {\tt DLAM} the following hold, for all $l, l' \in L$:
\begin{eqnarray}
 \bdia{A}(l \vee l' ) & =& \bdia{A}(l) \vee \bdia{A}(l')\\
 \Box_A (l \wedge l')  & =& \Box_A (l) \wedge \Box_A (l')\\
 \bdia{A}(l\wedge l') &\leq& \bdia{A}(l) \wedge \bdia{A}(l')\\
 \Box_A (l) \vee \Box_A (l') &\leq& \Box_A (l\vee l')\\
 \bdia{A}(\bot) = \bot &\quad&\Box_A(\top) = \top\\
 \bdia{A}(\Box_A (l)) &\leq& l\\
 l &\leq& \Box_A (\bdia{A}(l))
\end{eqnarray}
\end{proposition}
\bpf 
(4) follows from (1) and (3); similarly (5) follows from (2) and (3). (6) follows routinely from (1); similarly (7) follows from (1). (8) is routine, using (3), $\bot \leq \Box_A(\bot)$ and $ \bdia{A}(\top) \leq \top$. (9) follows from (3) and $\Box_A(l) \leq \Box_A(l)$; (10) is similar.  $\Box$

\bigskip
Let $L$ be a {\tt DLAM}. An {\em interpretation} of the set $M$ of formulae {\em in}  $L$ is a map $\semantics -\colon At \to L$. The meaning of formulae is obtained  by induction on the structure of the formulae:
\begin{eqnarray*}
\semantics {m_1 \vee m_2} = \semantics {m_1} \vee   \semantics {m_2}, &\quad& \semantics {m_1 \wedge m_2} = \semantics {m_1} \wedge   \semantics {m_2},\\
 \semantics {\bdia{A}(m)} = \bdia{A}(\semantics{m}), &\quad&  \semantics {\Box_A m} = \Box_A \semantics{m},\\
 \semantics{\top} = \top, &\quad& \semantics{\bot} = \bot\,.
\end{eqnarray*}
The  meanings of items and of contexts are obtained by mutual induction on their structure:
\begin{eqnarray*}
 \semantics{m} & =  &\text{as above}\\
 \semantics{\Gm^A} & = & \bdia{A}(\semantics {\Gm})\\
 \semantics {I_1, \cdots, I_n} & = & \semantics{I_1} \wedge \cdots \wedge \semantics{I_n}\\
 \semantics{\emptyset} & =  & \top
\end{eqnarray*}

Note that, since $\wedge$ is commutative and associative, the meaning of a context $\Gm$ is independent of its presentation as a list of items in a particular order.

A sequent $\Gm \vdash m$ is {\em true} in an interpretation $\semantics{-}$ in $L$ iff $\semantics{\Gm} \le \semantics {m}$; 
it is {\em true} in $L$ iff true in all interpretations in $L$, 
and it is {\em valid} iff true in every {\tt DLAM}. 
 
\begin{lemma}\label{inherit}
Let $\Gm, \Gm'$ be contexts with $\semantics{\Gm} \leq \semantics{\Gm'}$ and $\Dl$ a context-with-a-hole. Then $$\semantics{\Dl[\Gm]} \leq \semantics{\Dl[\Gm']}.$$ 
\end{lemma}
\bpf Routine induction on the structure of $\Dl$ (using also a similar result for items-with-a-hole). \qed
\begin{theorem} [Soundness]
Any derivable sequent is valid, 
	i.e.\  $\Gamma \vdash m$  implies $\semantics{\Gm} \leq \semantics m$ is {\em true} in any  interpretation $\semantics{-}$  in any  {\tt DLAM}.
\end{theorem}
\bpf We show that the axioms of the sequent calculus are valid and that the rules are truth-preserving. 
\begin{itemize}
\item Axioms. These are routine.
\item  The right rules. 
	\begin{itemize}
			\item $\wedge R$ and $\vee R$ are routine.
			\item $\bdia{A} R$.  We have  to show  
				\[
				\semantics{\Gm} \leq \semantics{m}  \quad \text{implies} 
			       \quad \semantics{\Gm', \Gm^A} \leq 			\semantics{\bdia{A}(m)}
				\]
					Assuming  $\semantics{\Gm} \leq \semantics{m} $, by definition of $\semantics{-}$ 
			 		we have to 		show 
			 		$\semantics{\Gm'} \wedge \bdia{A}(\semantics{\Gm}) \leq \bdia{A}(\semantics{m})$, 
			 		which follows by monotonicity of  $\bdia{A}$ and definition of meet.  
			\item  $\Box_A R$. We have to show
				\[
					\semantics{\Gm^A} \leq \semantics{m} \quad \text{implies} \quad \semantics{\Gm} \leq 					\semantics{\Box_A m}
				\]
			This follows directly from the definition of $\semantics{-}$ and  
			property (1) in the definition of a {\tt DLAM} as follows 
			\[
			\bdia{A}(\semantics{\Gm}) \leq \semantics{m} 
			\quad \text{iff} \quad \semantics{\Gm} \leq \Box_A 			\semantics{m}
			\]
	\end{itemize}
\item The left rules. These are done by induction on the structure of $\Dl$
	\begin{itemize}
		\item $\wedge L$ and $\vee L$ are routine.
		\item $\bdia{A} L$, we have to show
		\[
		\semantics{\Dl[m^A]} \leq \semantics{m'}\quad \text{implies} \quad  
		\semantics{\Dl[\bdia{A}(m)]} \leq \semantics{m'}
		\]
		which easily follows from the definition of $\semantics{\ }$.
		\item $\Box_A L$, we have to show 
		\[
		\semantics{\Dl [(\Box_A m, \Gm)^A, m]} \leq \semantics{m'} \quad \text{implies} \quad 
		\semantics{\Dl			[(\Box_A m, \Gm)^A]} \leq \semantics{m'}
		\]
		for which it is enough to show
		\[
		\semantics{\Dl[(\Box_A m, \Gm)^A]} \leq \semantics{\Dl [(\Box_A m, \Gm)^A, m]}
		\]
		By definition of contexts (and items) with holes this breaks down to three cases
		\begin{enumerate}
				\item $\semantics{(\Box_A m, \Gm)^A} \leq \semantics{(\Box_A m, \Gm)^A, m}$
						which  by definition of $\semantics{\ }$ is equivalent  to the following
						\[
						\bdia{A}(\Box_A \semantics{m} \wedge \semantics{\Gm}) 
							\leq \bdia{A}(\Box_A \semantics{m} \wedge \semantics{\Gm}) \wedge m
						\]
						and follows since by proposition~\ref{weakdist} and definitions 
						of $\bdia{A}$ and $\wedge$ we have
						\[
						\bdia{A}(\Box_A \semantics{m} \wedge \semantics{\Gm})  
						\leq \bdia{A}(\Box_A \semantics{m}) \wedge \bdia{A}(\semantics{\Gm})  
						\leq \bdia{A}(\Box_A \semantics{m}) \leq m
						\]
				\item  $\semantics{\Gm', J[(\Box_A m, \Gm)^A]} 
						\leq \semantics{\Gm', J[(\Box_A m, \Gm)^A, m]}$ 
						follows from case 1 by recursively unfolding 
						the definition of an item-with-a-hole.
				\item  $\semantics{\Dl'[(\Box_A m, \Gm)^A]^B} 
							\leq \semantics{\Dl'[(\Box_A m, \Gm)^A, m]^B}$ 
						follows from case 1 by recursively unfolding the 
						definitions of a context-with-a-hole and an item-with-a-hole.
		\end{enumerate}
	\end{itemize}
\end{itemize}
\qed		

\begin{theorem} {\bf Completeness.}
Any valid sequent is derivable, 
	i.e. if  $\semantics{\Gamma} \leq \semantics {m}$ for every {\tt DLAM} and every  interpretation  $\semantics{-}$ therein, then  $\Gamma \vdash m$.
\end{theorem}

\noindent
{\bf Proof}.  We follow   the Lindenbaum-Tarski proof method of completeness (building the counter-model) and show the following
\begin{enumerate}
\item  The logical equivalence $\cong$ defined as $\vdash \dashv$ over the formulae in $M$ is an equivalence relation, i.e. it is reflexive, transitive (by the admissibility of {\em Cut}), and symmetric.
\item The order relation $\leq$ defined as $\vdash$ on the above equivalence classes is a partial order, i.e. reflexive, transitive and anti-symmetric.
\item The operations $\wedge, \vee, \bdia{A}$, and $\Box_A$ on the above equivalence classes (defined in a routine fashion) are  well-defined.   To avoid confusion with the brackets of the sequents, i.e. $\Dl [\Gm']$, we occasionally drop the brackets of the equivalence classes and for example write $\bdia{A}(m)$ for $[\bdia{A} (m)]$. \begin{enumerate}
\item For $\bdia{A} [m] := [\bdia{A}(m)]$ we show 
\[
[m] \cong [m'] \implies [\bdia{A}(m)] \cong [\bdia{A}(m')]
\]
The proof tree of one direction is as follows, the other direction is identically easy
 \[
 \infer[\bdia{A} L]{\bdia{A}(m) \vdash \bdia{A}(m')}{\infer[\bdia{A} R]{m^A \vdash \bdia{A}(m')}{m \vdash m'}}
 \]
\item For $\Box_A [m] := [\Box_A m]$ we show
\[
[m] \cong [m'] \implies [\Box_A m] \cong [\Box_A m']
\]
The proof tree of one direction is as follows, the other direction is identically easy
\[
\infer[\Box_A R]{\Box_A m \vdash \Box_A m'}{\infer[\Box_A L]{(\Box_A m)^A \vdash m'}{m \vdash m'}}
\]
\item Similarly for $[m_1] \wedge [m_2] := [m_1 \wedge m_2]$ and $[m_1] \vee [m_2] := [m_1 \vee m_2]$.
\end{enumerate}
\item The above operations satisfy the properties of a distributive adjoint lattice (i.e. with binary joins and meets). 
\begin{enumerate}
\item  The proof tree for one direction of join preservation of $\bdia{A}$ is as follows, the other direction is also easy.
\[
\infer[\bdia{A} L]{\bdia{A} (m_1 \vee m_2) \vdash \bdia{A}(m_1) \vee \bdia{A}(m_2)}{
\infer[\vee L]{(m_1 \vee m_2)^A \vdash  \bdia{A}(m_1) \vee \bdia{A}(m_2)}{
	\infer[\vee R1]{m_1^A \vdash  \bdia{A}(m_1) \vee \bdia{A}(m_2)}{
		\infer[\bdia{A} R]{m_1^A \vdash \bdia{A}(m_1)}{\infer[Id]{m_1 \vdash m_1}{}}}
&
	\infer[\vee R2]{m_2^A \vdash  \bdia{A}(m_1) \vee \bdia{A}(m_2)}{
		\infer[\bdia{A} R]{m_2^A \vdash \bdia{A}(m_2)}{\infer[Id]{m_2 \vdash m_2}{}}}
}} 
\qquad 
\infer[]{}{}
\]
\item The proof trees for the adjunction between $\bdia{A}$ and $\Box_A$ are as follows
\[
\infer[\bdia{A} L]{\bdia{A}(m) \vdash m'}{
	\infer[\Box_A Inv]{m^A \vdash m'}{m \vdash \Box_A m'}}
\qquad \qquad
\infer[\Box_A R]{m \vdash \Box_A m'}{
	\infer[\bdia{A} Inv]{m^A \vdash m'}{\bdia{A}(m) \vdash m'}}
\]
\end{enumerate}
\end{enumerate}
\qed

\subsection{Examples of Algebraic Semantics}
We point out some examples for the algebraic semantics of our calculus. 

\begin{example}
The simplest example of a {\tt DLAM} is a Heyting Algebra:
\begin{proposition}\label{egHeyting}
A  Heyting Algebra $H$ is a {\tt DLAM} over an arbitrary  set $\cal A$. 
\end{proposition}
To see this let  $\bdia{A}(-)$ be $h \wedge -$ \  for some $h \in H$, then, since  $\wedge$ is residuated, the Galois right adjoint  to $\bdia{A}$ exists and is obtained  from  the implication. For instance we can set
 \begin{itemize}
\item $\bdia{A}(-) = \top \wedge -$ \ and we obtain  $\bdia{A} = \Box_A = id$, 
\item $\bdia{A}(-) = \bot \wedge - = \bot$ \ and we obtain  $\Box_A  = \top$,
\item $H =  {\cal A}$ \ and we obtain $\bdia{A}(-) = A \wedge -$,   hence $\Box_A - = A \supset -$ \ where $\supset$ is the implication. 
\end{itemize}
\end{example}

\begin{example}
One can argue that in a Heyting Algebra meets are commutative and idempotent but our $\bdia{A}$s generally are not. So a closer  match would be a residuated lattice monoid:
\begin{proposition}\label{egQuantal}
A residuated lattice monoid  $Q$ is a {\tt DLAM} over an arbitrary  set $\cal A$. 
\end{proposition}
 Recall that a residuated lattice monoid   $Q$ is  a lattice $(Q, \vee, \wedge, \top, \bot)$ with a monoid structure $(Q, \bullet, 1)$ such that the monoid multiplication preserves the joins and has a right adjoint in each argument, i.e. $q \bullet - \dashv -/q$ and $- \bullet q \dashv q\setminus -$. Thus if we take $\bdia{A}(-)$ to be either $q \bullet -$ or $- \bullet q$ then it will have a right adjoint in each case. 
For instance,  we can set
\begin{itemize}
\item   $\bdia{A}(-) = 1 \bullet -$  or $- \bullet 1$ and obtain  $\bdia{A} = \Box_A = id$, 
\item or set $\bdia{A}(-) = \bot \bullet -$ or $- \bullet \bot$ and obtain a   $\Box_A$ which is   a bi-negation operator, i.e. $\neg^l - = - / \bot$ and $\neg ^r - = \bot\setminus -$ respectively for each argument. \item Alternatively, we can have $L =  {\cal A}$ \ and thus obtain $\bdia{A}(-) = A \bullet -$,   hence $\Box_A - =  - / A$, and similarly for the other argument.
\end{itemize}
\end{example}

\subsection{Relational Semantics}
In this section, we  develop a Hilbert-style logic $APML_{Hilb}$ for our previous syntax and show that this logic provides an axiomatization for $APML_{Tree}$. We show that  this logic is sound and complete with regard to \emph{ordered Kripke Frames}, by applying the general Salqhvist  theorem for distributive modal logics, developed by Gehrke et al in~\cite{GehrkeNagahashiVenema}.

The set of formulae $M$  is the same as that of $APML_{Tree}$. Since the language does not include implication, following Dunn~\cite{Dunn},  the sequents are of the form $m \vdash m'$ for $m, m' \in M$.  The axioms and rules are:
\begin{center}
\fbox{\begin{minipage}{14cm}
{\bf Axioms.}
\[m \vdash m, \qquad \bot \vdash m, \qquad m \vdash \top
\]
\[
m \wedge (m' \vee m'') \vdash (m \wedge m') \vee (m \wedge m'')
\]
\[
m \vdash m \vee m', \qquad m' \vdash m\vee m', \qquad m \wedge m' \vdash m, \qquad m \wedge m' \vdash m'
\]
\[
\bdia{A}(m\vee m') \vdash \bdia{A}(m) \vee \bdia{A}(m'), \qquad \bdia{A}(\bot) \vdash \bot
\]
\[
\Box_A m \wedge \Box_A m' \vdash \Box_A (m \wedge m'), \qquad \top \vdash \Box_A \top
\]
\[
\bdia{A}(\Box_A m) \vdash m, \qquad m \vdash \Box_A \bdia{A}(m)
\]
{\bf Rules.}
\[
\infer[cut]{m \vdash m''}{m \vdash m' \quad m' \vdash m''}
\]
\[
\infer[\vee]{m \vee m' \vdash m''}{m \vdash m'' \quad m'\vdash m''} \qquad \infer[\wedge]{m \vdash m' \wedge m''}{m \vdash m' \quad m \vdash m''}
\]
\[
\infer[\bdia{A}]{\bdia{A}(m) \vdash \bdia{A}(m')}{m \vdash m'}\qquad \infer[\Box_A]{\Box_A m \vdash \Box_A m'}{m \vdash m'}
\]
\end{minipage}}
\end{center}

\begin{proposition} \label{HILB-soundcomplete}
$APML_{Hilb}$ is sound and complete with respect to DLAMs.
\end{proposition}

\bpf Soundness is easy. Completeness follows from a routine Lindenbaum-Tarski construction.
\endproof

\begin{proposition}\label{AML-GHV}
A sequent of the form $m \vdash m'$ is derivable in $APML_{Tree}$ if and only if it is derivable in $APML_{Hilb}$.
\end{proposition}
\bpf Follows from proposition ~\ref{HILB-soundcomplete}.\endproof

A Hilbert-style modal logic is \emph{Sahlqvist} whenever its modal axioms correspond to  first order conditions of a Kripke frame. According to Sahlqvist's Theorem, these modal logics are sound and complete with regard to their corresponding canonical Kripke models~\cite{Blackburn}. 

\begin{proposition}\label{Sahlqvist}
$APML_{Hilb}$ is Sahlqvist.
\end{proposition}

\bpf
It suffices to show that the two axioms  $m \vdash \Box_A \bdia{A}(m)$ and $\bdia{A}(\Box_A m) \vdash m$ are Sahlqvist. According to the method developed in~\cite{GehrkeNagahashiVenema}, the former  sequent is Sahlqvist if and only if $m$ is left Sahlqvist and $\Box_A \bdia{A}(m)$ is right Sahlqvist.  The former is obvious, the negative generation tree of the latter is as follows
\[
- \rTo (\Box_A, -) \rTo (\bdia{A}, -) \rTo (m, -)
\]
This is right Sahlqvist since the only choice node $\Box_A$ does not occurs in the scope of the only universal node $\bdia{A}$.   The proof of $\bdia{A}(\Box_A m) \vdash m$ being Sahlqvist is similar.
\endproof

For a Kripke semantics, we consider a simplification of that in~\cite{GehrkeNagahashiVenema}:
\begin{definition}\label{frame}
A {\em multi-modal Kripke frame} for $APML_{Hilb}$ is a tuple $(W, \leq, \{R_A\}_{\cal A}, \{R_A^{-1}\}_{A \in {\cal A}})$, where $W$ is a set of worlds, each $R_A$  is a binary relation on $W$ and $R_A^{-1}$ is its converse, and $\leq$ is a partial order on $W$ satisfying
\[
\leq  \circ  R_A^{-1}  \circ  \leq  \ \ \subseteq \ \ R_A^{-1} \qquad \text{and} \qquad
\geq  \circ  R_A        \circ  \geq  \ \   \subseteq \ \ R_A
\]
\end{definition}
A {\em Kripke structure} for $APML_{Hilb}$ is a pair  $\mathbb{M} = (\mathbb{F}, V)$  where $\mathbb{F}$ is a multi-modal Kripke frame for $APML_{Hilb}$  and  $V \subseteq W \times P$  is a valuation. Given such a  Kripke structure, a satisfaction relation  $\models$ is defined on $W$ and formulae of $APML_{Hilb}$ in the routine fashion. The clauses for the modalities are as follows:
\begin{itemize}
\item  $\mathbb{M}, w \models \bdia{A} (m)$ \quad iff \quad $\exists v \in W,  \quad w R_A^{-1} v \quad \ \mbox{and} \quad \ \mathbb{M}, v \models m$ 
\item  $\mathbb{M}, w \models \Box_A m$ \quad \ \ iff \quad $\forall v \in W, \quad  w R_{A}  v \quad \mbox{implies} \quad \mathbb{M}, v \models m$ 
\end{itemize}
From the general Sahlqvist theorem of~\cite{GehrkeNagahashiVenema} for distributive modal logics and our propositions~\ref{Sahlqvist} and~\ref{AML-GHV} it follows that
\begin{theorem}
$APML_{Tree}$ is sound and complete with respect to Kripke structures for  $APML_{Hilb}$.
\end{theorem}

\subsection{Representation Theorem}
 We end this section by stating some results about a concrete construction for DLAMs and a representation theorem for \emph{perfect} DLAMs. They follow from our previous results together with the general results of~\cite{GehrkeNagahashiVenema} about representation theorems for distributive modal logics; the definitions are from \cite{GehrkeNagahashiVenema}.
 
\begin{definition} The {\em complex} or {\em dual} algebra of a multi-modal Kripke frame for $APML_{Hilb}$ is the collection of the subsets of $W$ that are downward closed with respect to $\leq$. 
\end{definition}
\begin{definition}
A distributive lattice is called \emph{perfect}  whenever it is complete, completely distributive, and join generated by (i.e. each element of it is equal to the join of) the set of all of  its completely join irreducible elements. 
\end{definition}
\begin{lemma}
The complex algebra of a multi-modal Kripke frame for $APML_{Hilb}$ is closed under intersection, union and the modal operators $\Box_A Z := \{w \mid \forall v \in W,  wRv \implies v \in Z\}$ \ and \ $\Diamond_A Z = \{w \mid \exists v \in W,  wR^{-1}v, v \in Z\}$ for $Z \subseteq W$.
\end{lemma}
\begin{proposition}\label{complexDLAM}
The complex algebra of a multi-modal Kripke frame for $APML_{Hilb}$ is a perfect DLAM.
\end{proposition}
\begin{theorem}
Given a perfect DLAM $\cal L$, there is a frame whose complex algebra is isomorphic to $\cal L$.  
\end{theorem}

\bpf By the above proposition~\ref{complexDLAM}, it suffices to construct a frame from $\cal L$ in a way that the  complex algebra of the frame  is isomorphic to $\cal L$. As shown in lemma 2.26 and proposition 2.25 of~\cite{GehrkeNagahashiVenema}, the \emph{atom structure} of a  perfect DLAM is a  such a frame. \endproof

\section{Epistemic Applications}
Following previous work~\cite{BaltagCoeckeSadr,SadrThesis,SadrOckham}, we interpret $\bdia{A}(m)$ as `agent $A$'s uncertainty about $m$', that is, in effect, the conjunction of all the propositions that $A$ considers as possible when in reality $m$ holds. Accordingly, $\Box_A m$ will be interpreted `agent $A$ has information that $m$'. We could use the terminology of {\em belief}, but wish to avoid this as too suggestive about mental states. Agents can cheat and lie, so ``knowledge'' is inappropriate.

The intended application of our calculus is scenarios where extra information is available about the uncertainty of agents. This will always be of the form of one or more assumptions of the form 
$\bdia{A}(p) \supset m''$ where $p$ is an atom and $m''$ is a disjunction of atoms, e.g. $p_1 \vee p_2$. Such assumptions express ideas that would, in the relational semantics, be encoded in the accessibility relation, e.g.\ that such and such a world can access certain other worlds. Such implications are not even formulae of our language; we can however add them as follows, by adding (for each such given assumption) the following evidently sound rule
\[
\infer[Assn]{\Delta[(\Gm, p)^A] \vdash m}{\Delta [(\Gm,p)^A, m''] \vdash m}\qquad \qquad 
\]

It is routine to note that the proofs of admissibility of {\em Weakening} (\ref{weak-lab}) and of {\em Contraction} (\ref{Icontr}) still work when these extra rules are considered; it is important for example that the principal item of $Assn$ be of the form $(\Gm, p)^A$ rather than $\bdia{A}(p)$. The same applies to the invertibility lemmas.

\begin{proposition}
The {\em Cut} rule is admissible in $APML_{Tree}^{Assn}$.
\end{proposition}
\bpf
There are three extra cases: 
\begin{enumerate}
\item[(xii)] The first premiss is an instance of $Assn$:
$$
\infer[Cut]{\Delta' [\Delta [(\Gm, p)^A]] \vdash m'}
	{\infer[Assn]
		{\Delta[(\Gm, p)^A] \vdash m}{\Delta[(\Gm, p)^A, m''] \vdash m}
		&
		\Delta'[m] \vdash m'
		}
$$
is transformed to
$$
\infer[Assn\ .]
	{\Delta'[\Delta[(\Gm, p)^A]] \vdash m'}
		{\infer[Cut]
			{\Delta'[\Delta[(\Gm, p)^A, m'']] \vdash m'}
				{\Delta[(\Gm, p)^A, m''] \vdash m 
				\qquad 
				\Delta'[m] \vdash m'
				}
		}
$$
\item[ (xi)(m)]  The first premiss is an instance of $\Box_A R$ and the second premiss is an instance of $Assn$, with  the cut formula $\Box_A m$  non-principal, i.e.\ not occurring as an element in the principal item of $Assn$.
$$
\infer[Cut]
	{\Delta[\Gamma][(\Gm', p)^B] \vdash m'}
		{\infer[\Box_A R]
			{\Gamma \vdash \Box_A m}
				{\Gamma^A \vdash m}
		&
		\infer[Assn]
			{\Delta[\Box_A m][(\Gm', p)^B] \vdash m'}
				{\Delta[\Box_A m][(\Gm', p)^B, m''] \vdash m'}
		}
$$
is transformed to
$$
\infer[Assn]
	{\Delta[\Gamma] [(\Gm', p)^B] \vdash m'}
		{\infer[Cut]
			{\Delta[\Gamma][(\Gm', p)^B, m''] \vdash m'}
				{\infer[\Box_A R]{\Gamma \vdash \Box_A m}{\Gamma^A \vdash m}
				&
				\Delta[\Box_A m][(\Gm', p)^B, m''] \vdash m'
				}
}
$$

\item[(xi)(n)]  The first premiss is an instance of $\Box_A R$ and the second premiss is an instance of $Assn$, with  the cut formula $\Box_A m$ principal, i.e.\ occurring in the principal item of $Assn$.

$$\infer[Cut]
	{\Delta[(\Gm', \Gm, p)^B] \vdash m'}
		{\infer[\Box_A R]
			{\Gamma \vdash \Box_A m}
				{\Gamma^A \vdash m}
		&
		\infer[Assn]
			{\Delta[(\Gm', \Box_A m, p)^B] \vdash m'}
				{\Delta[(\Gm', \Box_A m, p)^B, m''] \vdash m'}
		}
$$
is transformed to
$$
\infer[Assn]
	{\Delta[(\Gm',\Gm, p)^B] \vdash m'}
		{\infer[Cut]
			{\Delta[(\Gm', \Gm, p)^B, m''] \vdash m'}
				{\infer[\Box_A R]{\Gamma \vdash \Box_A m}{\Gamma^A \vdash m}
				&
				\Delta[(\Gm', \Box_A m, p)^B, m''] \vdash m'
				}
}
$$
\end{enumerate}

\endproof

As an example consider the muddy children puzzle. It goes as follows:  $n$ children are playing in the mud and $k$ of them have  muddy foreheads. Each child can see the other children's foreheads,  but cannot see his own. Their father announces to them ``At least one of you has a muddy forehead.''  and then asks them ``Do you know it is you who has a muddy forehead?''.  After $k-1$ rounds of no answers by all the children, the muddy ones know that they  are muddy. After they announce it in a round of yes answers, the clean children know that they are not muddy. 

\newcommand{\overk}{\overline{k}}

To formalize this scenario, assume the children are enumerated and the first $k$ ones are muddy. Consider  the  propositional atoms $s_\beta$ for $\beta \subseteq  \{1,\cdots, n\}$ where  $s_\beta$ stands for the proposition that exactly the children in $\beta$ are muddy and $s_\emptyset$ stands for `no child is muddy'. The formula $\bdia{i}(s_\beta)$  stands for  the uncertainty of child $i$ about each of these atoms  before father's announcement. Since child $i$ can only see the other children's foreheads and not his own, he is uncertain about himself being muddy or not. 
 Let $\overk$ = the set $\{1,\dots,k\}$ (we write $s_{1, \cdots, k}$ rather than $s_{\{1, \cdots, k\}}$), then the assumption for the  uncertainty of  the muddy child $i$ is  $\bdia{i}(s_{\overk}) \supset s_{\overk} \vee s_{\overk \setminus i}$, captured in the calculus by the following instance of the assumption  rule
\[
\infer[Assn]
	{\Dl[(\Gm, s_{\overk})^i] \vdash m}
		{\Dl[(\Gm, s_{\overk})^i, s_{\overk} \vee s_{\overk \setminus i}] \vdash m}
\]
The assumption for the uncertainty of the clean child $w$ is   $\bdia{w}(s_{\overk}) \supset s_{\overk} \vee s_{\overk \cup w}$ and its $Assn$ rule is similar.  For $1 \leq i,j \leq k$  and $k+1 \leq w$
we have that before the $k-1$'th announcement a muddy  child $i$ is uncertain about having a muddy  forehead: $s_{\overk} \vdash \Box_i (s_{\overk} \vee s_{\overk \setminus i})$ 
(i.e.\ $\bdia{i}(s_{\overk}) \vdash s_{\overk} \vee s_{\overk \setminus i}$). The proof tree of this property is as follows
\[
	\infer[\Box_i R]
		{s_{\overk} \vdash \Box_i (s_{\overk} \vee s_{\overk \setminus i})}
					{\infer[Assn]
						{(s_{\overk})^i \vdash s_{\overk} \vee s_{\overk \setminus i}}
							{\infer[Id]
								{(s_{\overk})^i, s_{\overk} \vee s_{\overk \setminus i} 
							 	 \vdash s_{\overk} \vee s_{\overk \setminus i}}
							  	{}
							}
					}
\]
The uncertainties of children change after each announcement as follows\footnote{The way the uncertainties change after each announcement is formalized in the sequent calculus of previous work~\cite{BaltagCoeckeSadr,SadrThesis} via adding a dynamic logic for actions and extra rules for  epistemic update; that calculus was not cut-free; Here, we change the assumptions of these uncertainties by hand and defer a full formalization to future work, for more details see next section.}: the $k$'s announcements eliminates the $s_\gamma$ disjunct from the uncertainty before the announcement when  $\gamma \subseteq  \{1,\cdots, n\}$  is of size $k$ ; father's announcement eliminates the  $s_\emptyset$ disjunct.  For example, after the series of 1 to $k-1$'th announcements all the disjuncts except for $s_{\overk}$ will be eliminated from muddy child $i$'s uncertainty, hence his previous uncertainty assumption rule changes to
\[
\infer[Assn]
	{\Dl[(\Gm, s_{\overk})^i] \vdash m}
		{\Dl[(\Gm, s_{\overk})^i, s_{\overk}] \vdash m}
\]
 The assumption for the uncertainty of the clean child $w$ changes in a similar way.
We have that, after the $k-1$'th announcement, a muddy child $i$ obtains information (a) that he is muddy and (b) that other muddy children also  obtain information that he is muddy: 
$s_{\overk} \vdash (\Box_i s_{\overk}) \wedge (\Box_i \Box_j s_{\overk})$. However, a clean child $w$ will be uncertain about being muddy before and after the $k-1$'th announcements:  $s_{\overk} \vdash \Box_w (s_{\overk} \vee s_{\overk \cup w})$. 
The proof tree of the property for a muddy child $i$ (where child $j$ is also muddy) is as follows: 
$$
\infer[\wedge R]
	{s_{\overk}  \vdash (\Box_i s_{\overk})  \wedge (\Box_i \Box_j s_{\overk}) }
		{\infer[\Box_i R]
			{s_{\overk}  \vdash \Box_i s_{\overk} }
				{\infer[Assn]
					{(s_{\overk})^i \vdash s_{\overk} }
						{\infer[Id]
							{(s_{\overk})^i, s_{\overk}  \vdash s_{\overk} }
								{}
						}
				}
 		& 
 		\infer[\Box_i R]
			{s_{\overk}  \vdash \Box_i \Box_j s_{\overk} }
				{\infer[Assn]
					{(s_{\overk})^i \vdash \Box_j s_{\overk} }
						{\infer[\Box_j R]
							{(s_{\overk})^i, s_{\overk}  \vdash \Box_j s_{\overk} }
									{\infer[Assn]
										{((s_{\overk})^i, s_{\overk})^j \vdash s_{\overk}}
											{\infer[Id]
												{((s_{\overk})^i, s_{\overk})^j, s_{\overk} \vdash s_{\overk}}
													{}
											}
									}
{}}}}
$$
Consider a twist to the above scenario. Suppose that none of the children are muddy but that the father is a liar  (or he cannot see properly) and the children do not suspect this (thus their uncertainties change in the same way as above). After father's false announcement, any child $i$ will (by reasoning) obtain false information  that he is the only muddy child: $s_\emptyset \vdash \Box_i s_i$.  
The proof tree is as follows:  
$$ 
	\infer[\Box_i R]
		{s_\emptyset \vdash \Box_i s_i}
				{\infer[Assn]
					{(s_\emptyset) ^i \vdash  s_i}
						{\infer[Id]
							{(s_\emptyset) ^i, s_i \vdash s_i}
								{}
						}
				}
$$

\section{Conclusion and Future Work}
We have developed a tree-style sequent calculus for a positive modal logic where the modalities are adjoints rather than De Morgan duals.  We have shown that  the structural rules of {\em Weakening}, {\em Contraction} and {\em Cut} are  admissible in our calculus. We have also shown that our calculus is sound and complete with regard to  bounded distributive lattices with a pair of adjoint operators. Examples of these are complete Heyting Algebras and residuated lattice monoids. Using general results of~\cite{GehrkeNagahashiVenema}, we have shown that our  calculus  is sound and complete with respect to ordered Kripke frames, through developing a Hilbert-style calculus that has the same deductive power. We have motivated the applicability of our modal logic by encoding in it partial assumptions of real-life scenarios and  proving epistemic properties of agents in the  milestone puzzle of muddy children, but also to  a newer mis-information version of it where father's announcement is not necessarily truthful.  Since our box modality is not necessarily truthful, we can as well reason about the settings where agents obtain false information as a result of dishonest announcements. Our proof method of unfolding the adjunction and then renaming the left adjoint to its assumed values has made our proofs considerably simpler than the usual proof method of epistemic logics for  the muddy children puzzle which uses the fixed point operator of the box. Our calculus has been implemented \cite{Kriener} in Prolog; see this report for details of the (routine) decidability proof.  

Our logic may be seen as a positive version of  $K_t$, i.e.\ tense logic. Thus one can deduce that a proof theory thereof can be obtained by restricting any proof theory of tense logic to rules for  conjunction and disjunction, the existential past and the universal future that only satisfy the $K$ axiom. We are unaware of a tree-style proof theory for this kind of modal logic in the literature, noting that the presence of $T, 4, 5$ axioms make the proof theory far easier than their absence in the logic. Thus we believe that  our  tree-style deep inference proof theory and its automated decision procedure is novel and so is its application to epistemic scenarios.

Future directions of our work include:
\begin{itemize}
\item A tree-style  cut-free sequent calculus that is sound and complete with regard to a residuated monoid with adjoint modalities has been developed  in~\cite{Moortgat}, it can easily be extended to a Quantale.
We believe that pairing this calculus with what we have in this paper, that is  adding  to it the rules for the action of the Quantale on its right module such that it remains cut-free, will provide a cut-free sequent calculus. From this one obtains a  decision procedure for a distributed version of Epistemic Systems of~\cite{BaltagCoeckeSadr} and thus a negation-free version of Dynamic Epistemic Logic of~\cite{BaltagMoss}. This calculus will be an improvement on the algebraic decision procedure of~\cite{SimonSadr}, which only implements a sub-algebra of the algebra of Epistemic Systems (namely one that allows $\bdia{A}$ only on the right and $\Box_A$ only on the left hand side of the partial order).
\item A representation theorem for \emph{perfect} DLAMs follows from general results of~\cite{GehrkeNagahashiVenema}. But DLAMs need not be complete and completion involves introduction of, in principle,  infinitary lattice operations. In~\cite{CelaniJansana} similar results are obtained for positive modal logics where $\Box$ and $\Diamond$ come from the same relation; it might be possible to alter their duality theorem and make it suitable for our adjoint modal logic. However, those results, similar to that of~\cite{GehrkeNagahashiVenema}, are with respect to the less intuitive ordered frames. At the moment, we are more inclined towards working with the usual non-ordered frames, along the lines of~\cite{Dunn}, that is by using theory and counter-theory pairs to build our canonical frames.
\item  As shown in propositions~\ref{egHeyting} and~\ref{egQuantal}, Heyting algebras and residuated lattice monoids are examples of DLAMs. So in principle our nested tree sequents might be adapted to provide a new sound and complete cut-free  proof  system for the logics based on these algebras, i.e.\ for intuitionistic and linear logics where the conjunction and tensor (respectively) are treated as adjoint operators.   In the former case, we will need extra rules to take care of the commutativity of conjunction, but in the latter case we hope to obtain a new cut-free proof theory for non-commutative intuitionistic multiplicative linear logic. It is also worth investigating how logics with classical negation and thus de Morgan dual connectives can be formulated in this context.
\end{itemize}


\end{document}